# Design and characterization of high energy micro-CT with a laser-based X-ray source

Yue Yang[1], Yu-Chi Wu[1,2], Liang Li[3], Si-Yuan Zhang[3], Ke-Gong Dong[1], Tian-Kui Zhang[1], Ming-Hai Yu[1], Xiao-Hui Zhang[1,3], Bin Zhu[1], Fang Tan[1], Yong-Hong Yan[1], Gang Li[1], Wei Fan[1], Feng Lu[1], Zong-Qing Zhao[1,2,*], Wei-Min Zhou[1,2], Lei-Feng Cao[1,2] and Yu-Qiu Gu[1,2,*]

[1] Science and Technology on Plasma Physics Laboratory, Laser Fusion Research Center, CAEP, Mianyang 621900, Sichuan, China

[2] IFSA Collaborative Innovation Center, Shanghai Jiao Tong University, Shanghai 200240, China

[3] Department of Engineering Physics, Tsinghua University, Beijing 100084, China

* Corresponding author: yqgu@caep.cn and zhaozongqing99@caep.cn

**Abstract**

The increasingly demand for machining accuracy and product quality excites a great interest in high-resolution non-destructive testing (NDT) methods, but spatial resolution of conventional high-energy computed tomography (CT) is limited to sub-millimeter because of large X-ray spot size. Therefore, we propose a novel high-resolution high-energy CT based on laser-driven X-ray source and prove its feasibility to allow high-spatial-resolution tomographic imaging of dense objects. A numerical model is developed with a consideration of realistic factors including parameter fluctuations, statistical noise and detecting efficiency. By using modulation transfer functions, the system performance is quantitatively characterized and optimized in terms of source characteristics, detector aperture, geometrical configuration and projection parameters. As a result, the simulated tomography for a high-density object (up to 19.35g/cm$^3$) achieves a basic spatial resolution of 64.9μm. This concept expands the prospects of laser-based compact X-ray sources and shows a great potential to achieve high-perspectivity micro-CT imaging for various industrial applications.

**Key words:** high spatial resolution; high-energy CT; laser-driven X-ray source; micro spot; CT imaging

## 1. Introduction

X-ray imaging is a widely used method to probe and identify the construction of complex objects. To examine the product accuracy nondestructively, X-ray computed tomography (CT) [1] was invented to produce densitometric (that is, density and geometry) images of a cross-sectional plane through an object, thus permitting quantitative physical characterization of its internal structure [2]. Therefore, as an effective and intuitive test method, X-ray CT technology has been playing an important role in clinical medicine, non-destructive testing (NDT), manufacturing production, scientific research and national security [3].

The spatial resolution is a crucial indicator of the CT system capability, which strongly depends on the radiation spot size. For low-energy medical scanning and biomedical engineering, the advanced X-ray machines can provide a compact spot to achieve the so-called micro-CT with spatial resolutions up to tens of microns [4,5,6]. On the other hand, the hard X-rays for high-energy (multi-MeV or much higher) CT applied in industrial detection come from the bremsstrahlung radiation of fast electrons interacting with high-Z material targets [3,7]. However, conventional linear accelerators generally produce energetic electrons with millimeter-scale spot sizes, limiting the improvement on the spatial resolution [8,9].

As the technological application and innovation in manufacturing industries develops rapidly, the standard of industrial production has become increasingly important [10]. Faced with the approaching demand for high-spatial-resolution high-penetration tomography, the introduction of micro-spot high-energy brilliant X-ray source is urgently desired.

In recent years, great advances in laser technology have motivated numerous studies of laser-plasma X-ray source [11,12,13,14,15,16]. And laser wakefied acceleration [17,18] (LWFA) becomes a potential way to generate micro-spot bremsstrahlung source attributed to the production of collimated high-energy electrons with compact spot. The experiment performed at Laboratoire d'Optique Appliqué´e in 2005 first demonstrated a radiograohy of complex and dense objects with submillimeter resolution using γ-ray source from laser wakefield accelerators [11]. Then in 2011, an optimized spatial resolution of 30μm was reached with an average electron beam divergence of 3mrad (FWHM) and a quasi-monoenergetic energy spectrum (80MeV peak energy) [12]. Their recent work showed an improved electron charge of 1nC per shot with 1.1J laser pulse, making this source well-suited for high-yield X-ray radiography [16]. Our group has started researches on this subject since 2012 [19]. In 2016, we have obtained a hard X-ray source with an optimal spot size up to 40μm, and the spatial resolution better than 2.5LP/mm can be achieved in 2D radiographical demonstrations [20]. Therefore, such laser-driven X-ray source is able to reach much smaller spot sizes than conventional techniques do, bringing a chance to break the bottleneck of current high-energy CT resolution as well as to promote the development of other relevant technologies and applications.

In order to practically utilize this radiation source in CT technique, the feasibility analysis and conceptual design are necessary, and the systematic optimization is helpful to make the most of its advantages and avoid adverse effects. Additionally, the influence of the radiation instabilities on the system capability is an important factor which has not been investigated yet. So in this paper, we propose a laser-driven X-ray CT technology. Based on realistic data and numerical simulations, we demonstrate that high-spatial-resolution CT imaging can be successfully realized using laser-based, compact bremsstrahlung source. A realizable and practicable optimization to improve spatial resolution and contrast to noise ratio of CT images is obtained via parametric study. And the admissible value of spot sizes and fluctuations are discussed to reach an acceptable resolution. Consequently, our concept shows its feasibility to achieve high-spatial-resolution high-perspectivity tomography under actual conditions, offering design guide to the pragmatic CT devices.

This paper is organized as follows. Sec. II presents the basic tomographic theory and the method of our CT simulation. The initial system configuration in the numerical simulation is defined and confirmed. In Sec. III, we quantitatively discuss the dependences of the performance characteristics on different system parameters. Finally, the summary and conclusion are given in Sec. IV.

2. **Numerical simulation and evaluation method**

As we know, the CT image is a quantitative cross-sectional map revealing the spatial distribution of linear X-ray attenuation coefficient in the plane [21]. The linear attenuation coefficient $\mu$ is the rate at which the object material attenuates an X-ray with a particular spectrum during the scan. That makes it possible to use emulational computations based on numerical models to assess the potential of a specific CT system. In order to optimize the system design and acquire a high-quality image, we develop a theoretical simulation method which predicts the imaging result and foresees the system performance. In the following, a special system dependent on the laser-driven bremsstrahlung source is defined, upon which computational simulations of CT imaging are in detail performed to validate our proposal.

Primarily, the ranges of systematic parameters are analyzed to ideally reach our basic requirement on the spatial resolution. For a given CT system, the equivalent beam width (BW) determines the ultimate resolution [21]. BW (the image unsharpness) is a function of the detector aperture $d$ (the inherent detector unsharpness), the source width $a$ (the focal spot unsharpness), and the geometrical magnification factor $M$:

$$BW = \frac{\sqrt{d^2 + \left[a(M-1)\right]^2}}{M} \quad (1)$$

So the resolution strongly depends on the detector width, the spot size, and the geometric configuration. Fig. 1(a) and (b) present the dependencies of the theoretical resolution on the magnification factor, the spot size and the detector width. We can see that resolution is positively correlated to the spot size. For a fixed detector width of 100μm and a spot radius larger than 100μm, the spatial resolution can reach a minimum when the magnification factor increases from 1 to 1.8. Moreover, the resolution is found to depend both on the detector aperture and the magnification factor. Therefore, with an overall consideration of the theoretical analysis and experimentally accessible conditions, we set the detector width to be 100μm, the source size ranging from 100μm to 200μm, and the magnification factor within 1.1 to 1.6 to get an acceptable resolution below 100μm.

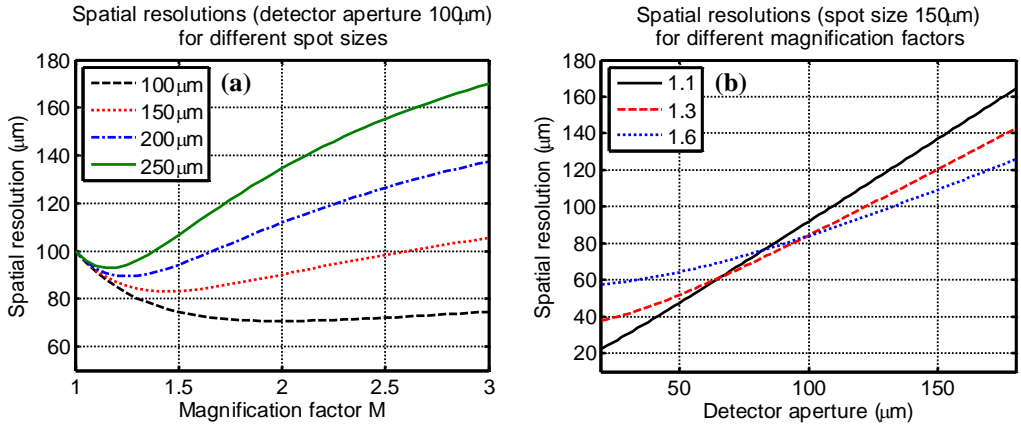

FIG. 1. Spatial resolutions calculated from Eq. 1: (a) detector aperture 100μm for different spot sizes and magnification factors; (b) spot size 150μm for different magnification factors and detector apertures.

Based on the theoretical design, the simulation CT system is developed and modeled by a self-made program neglecting the contribution of scattered radiation. Considering the realistic X-ray beam from laser-plasma interaction, incident photons are carried with energies between 0.05~10MeV, and the spectrum resembles as the typical distribution of bremsstrahlung from the experimental measurements (the fitting temperature is between 5 MeV and 8 MeV according to our recent experimental results in Ref. 20). The planar intensity of source is set as Gaussian distribution with the full width at half maximum (FWHM) of 150μm. It is noted that we didn't use the minimum size obtained in the experiment to keep a balance of the spot size and the photon yield. The continuous spectrum and planar distribution of the source are shown in Fig. 2(a) and (b).

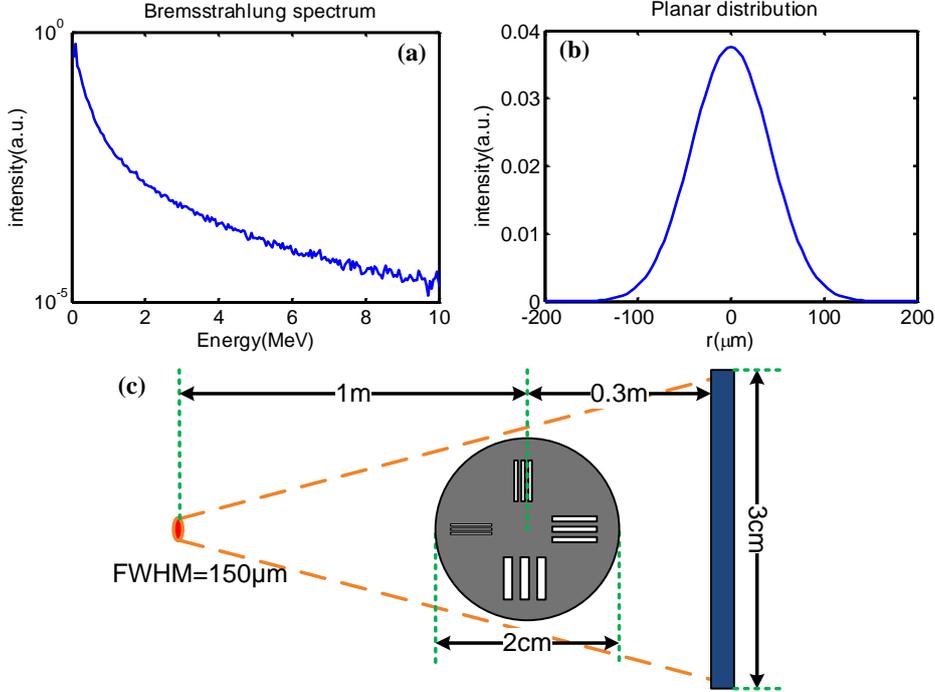

FIG. 2. Simulation settings: the (a) spectrum and (b) planar distribution of X-ray source; (c) the geometric setup.

In the simulation, the geometric setup is arranged as displayed in Fig. 2(c), the object-to-source distance is 1m and the image-to-source distance is 1.3m, corresponding to a magnification factor of 1.3. The detector material is chosen to be $CdWO_4$ which gives an absorption efficiency of 25% in 1MeV-10MeV to meet the detecting demand [9]. According to realistic commercial products, the detector array is chosen to be 3cm in length and 10mm in thickness, containing 300 linear units with a size of 100μm. To quantify the spatial resolution, the object is designed based on the existing resolution evaluation phantom [8] which has four groups of same-size gaps distributed at different directions respectively (see Fig. 2(c)), implicating different numbers of line pairs (3.3LP/mm to 10LP/mm). This object is chosen to be an iron ($7.86g/cm^3$) cylinder with a radius of 1cm.

Besides, the realistic situation is essentially influential in the system performance, so we introduced the Poisson distributed noise to reckon in the influence of X-ray scattering [22], statistical fluctuations and background noises. The absorption efficiency of detector is also taken into account in our computations. Specifically, the uncertainty in X-ray generation from laser-plasma interactions determines the practicality of this proposal. For instance, the fluctuation of electron charge leads to variant X-ray doses. The electron pointing instability and laser angle drift make changes of source position. And in actual measurement, the shot-to-shot position drift causes varied sizes of equivalent photon sources during multiple-shot accumulation for each digital radiograph. Therefore, the random fluctuation of source brilliance (±20%), position (±20μm) and size (±20μm) is assumed based on previous experimental results and system properties.

In the fan-beam projection procedure, a series of projection views are collected by the detector from X-ray scanning in multi-direction through a certain object plane. For every slice, the X-ray attenuation measurements are made along a set of paths projected at different locations around the periphery of the object. Therefore, the projection data is created by computing the line integral of a phantom, which can be expressed as

$$P(l) = -\ln \int dE s(E) e^{-\int_l \mu(E, \vec{r}) d\vec{r}} \tag{2}$$

where the weighting function $s(E)$ represents the X-ray spectrum for photon energy $E$, $r$ is a vector in the Euclidean space and $l$ denotes the ray path. It enables us to characterize different geometries and conditions by parameters. In the simulation, we get one slice of digital radiography (DR) every 0.5 degree, summed into 720 angles over 360 degrees. The photon number for every DR image is about $2.8 \times 10^8$ in this system. This parameter is appropriately chosen to have a fine signal-to-noise ratio, which will be later detailed in Sec. III. Thereupon, the obtained sinogram (the two-dimensional X-ray opacity measurements consisting of all the projections) is shown in Fig. 3(a). By using filtered backprojection (FBP) reconstruction technique [23] based on Central-Section Theorem [24], we are able to compute a tomographic image from the sinogram. The acquired views are convolved using Fourier transforms to apply a mathematical filter proportional to frequency. In this way, the description of the object is presented, indicating the distribution of radiation attenuation coefficients representing different materials on the projected cross-section.

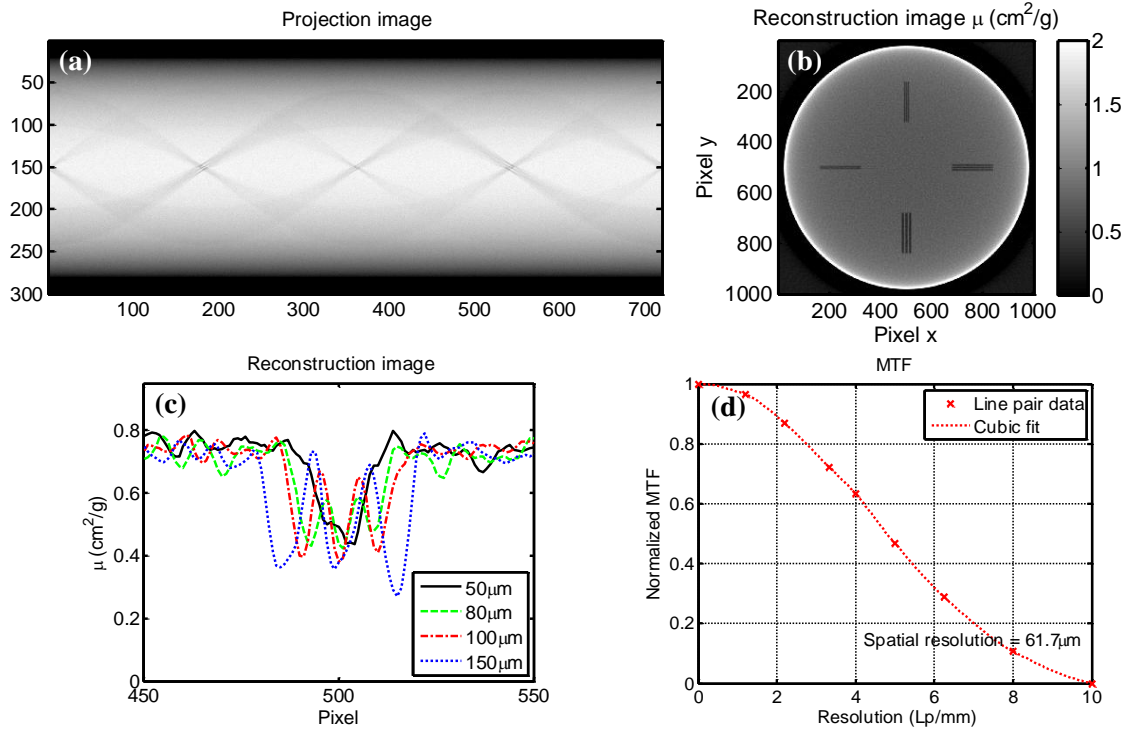

FIG. 3. Reconstructed results: (a) the sinogram (position within view vs. view angle); (b) CT image; (c)

the central profiles along the vertical line of each groups of gaps in (b); (d) MTF image. Simulation parameters: the FWHM of source: 150μm, the detector pixel size: 100μm, the geometrical magnification factor: 1.3; $2.8 \times 10^8$ photons/DR, 720 projection angles.

As a result, the reconstructed image consists of 1000×1000 pixels with a size of 21μm is presented in Fig. 3(b), while Fig. 3(c) shows the vertical central profiles. The construction of the object is clearly seen from the light-dark distribution of attenuation value, proving the accuracy of modeling and computation. And we roughly examine the resolution by distinguishing the adjacent gaps apart. Moreover, due to the polychromatic radiation, the beam hardening effect [25] causes a false radial density gradient in the image, resulting in low values at the center and high values at the periphery of the object observed in Fig. 3(b).

As an important system-level performance factor, basic spatial resolution is quantified as the smallest spacing at which two parts can be distinguished as separate entities. And the resolution of a CT system is generally measured from line patterns or modulation transfer function (MTF) [9]. In the present work, we calculate its value as below [8]: set the broadest rectangle platform contrast $\Delta\mu$ as basis, normalize the contrast $(\Delta\mu)_e$ of other line pairs to get the modulation data corresponding to different relative frequency response, then the spatial frequency of 10% modulation degree is the resolution of the system. Here, several objects with different line pairs are used to measure a series of modulation values. From the MTF curve (see Fig. 3(d)) fitted from line pattern data we get a spatial resolution of about 61.7μm, implying a good imaging quality. On the other hand, the factor CNR (contrast to noise ratio) [26,27] is used to evaluate the contrast resolution, which also affects the system performance under some circumstances, expressed as

$$CNR = \frac{E(\mu_{bg}) - \min(\mu_{signal})}{std(\mu_{bg})}$$

(3)

It is defined to be the difference between the ROI (region of interest) and the background region values of the optical properties, divided by the noise (the standard deviation) of background. Thereupon, the calculated CNR of the reconstructed image is 12.7. These results indicate the effectiveness of the designed scheme to theoretically achieve high-spatial-resolution high-perspectivity CT.

## 3. Discussion
### 3.1 Parameter optimization

According to our analysis, the decisive parameters for the tomographic process and the imaging quality are the detector aperture, the source size and the geometrical magnification factor. Also, the radiation intensity and projection angle in actual measurements are necessary to be included. In this section, we change these parameters to compare with the theoretical results and study the performance characteristics of this system. The influence of radiation instabilities on the CT test will be further discussed to give the prerequisite for a 100μm-resolution with laser-based X-ray sources.

Firstly, the roles of total photon number and projection angle are studied to get the demand of realistic scanning and accumulation. Fig. 4(a) and (b) shows the simulated CT images (central profiles) of photon number $2.8 \times 10^7$ and $1.4 \times 10^8$ every DR image respectively. The corresponding MTF image is accordingly computed and displayed in Fig. 4(c). In similar matter, results of different projection angle numbers are shown in Fig. 4(d), (e) and (f). It is found that larger radiation intensity results in better image quality, attributed to the decrease of statistical noise with respect to the signal. Therefore, accumulating laser shots to increase the total photon number per DR image is helpful to improve the performance in practical measurement. On the other hand, insufficient projection angles may cause the reconstructed defect which will lead to worse resolutions. The spatial resolutions derived from MTF images are listed in Table 1. Inferred from the results, over $1.4 \times 10^8$ photons/DR and 720 projection angles are sufficient to acquire a good resolution and tolerable noise level.

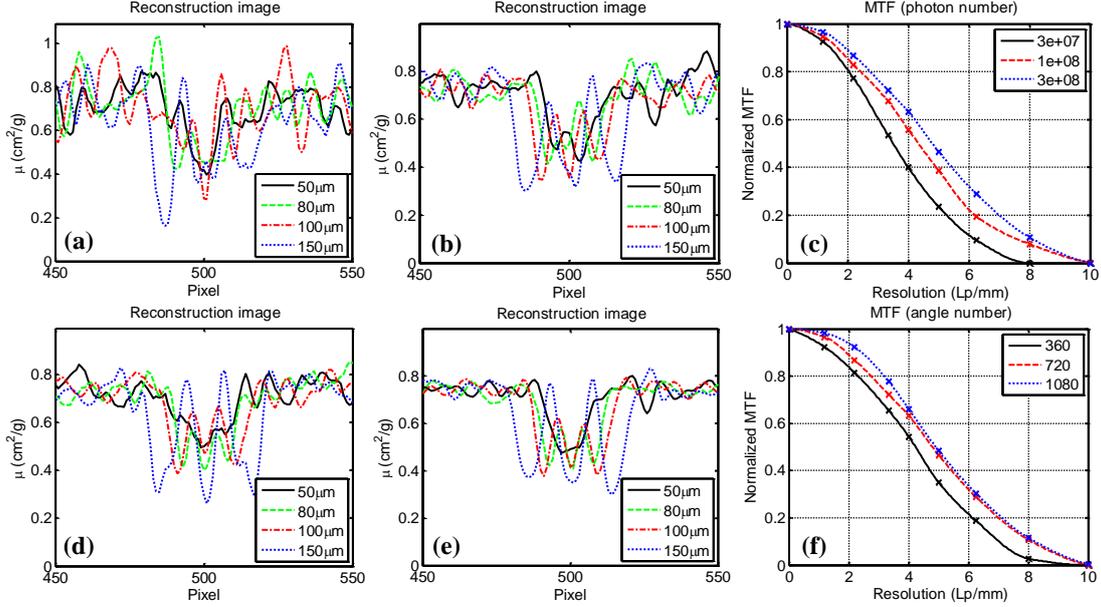

FIG. 4. Reconstructed results (central profiles) of different photon numbers: (a) $2.8 \times 10^7$ photons/DR; (b) $1.4 \times 10^8$ photons/DR; (c) MTF image. Central profiles of different projection angles: (d) 360; (e) 1080; (f) MTF image.

TABLE 1. Spatial resolutions for different parameters

| Different parameters | Number of photons/DR | | | Projection angle number | | | Geometrical magnification factor | | | Detector pixel size (μm) | | |
|---|---|---|---|---|---|---|---|---|---|---|---|---|
| | $2.8 \times 10^7$ | $1.4 \times 10^8$ | $2.8 \times 10^8$ | 360 | 720 | 1080 | 1.6 | 1.3 | 1.1 | 50 | 100 | 150 |
| Spatial resolution (μm) | 80.7 | 65.8 | 61.7 | 71.4 | 61.7 | 61.0 | 62.5 | 61.7 | 70.9 | 26.6 | 61.7 | 101.0 |

Next, the optimization of geometrical magnification factor is carried out by altering the object-to-source distance. The distances are chosen to be 0.5m, 1m and 3m while the detector keeps 0.3m from the object, corresponding to magnification factors of 1.6, 1.3 and 1.1. The reconstructed images are demonstrated in Fig. 5(a) and (b). One can infer that appropriately increasing the magnification factor is able to improve the imaging capabilities thanks to enhanced radiation intensity, but too large factor might impair the spatial resolution because of the concomitant growing effects of finite source size and non-negligible noises. Moreover, certain detector aperture also limits the further improvement of system resolution. Here we obtain a magnification factor of 1.3 with an object-to-source distance of 1m to get the optimal resolution, consistent with the outcome of Fig. 1. And it is noted that the simulated resolutions are generally less than the calculation from Eq. 1, due to the different definitions of system resolutions between the MTF evaluation (spatial frequency of 10% modulation degree) and the theoretical formula (equivalent beam width).

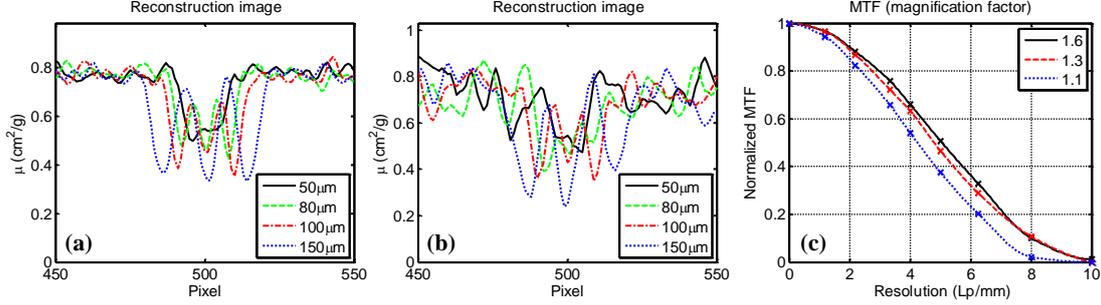

FIG. 5. Reconstructed results (central profiles) of different geometrical magnification factors: (a) 1.6; (b) 1.1; (c) MTF image.

Since larger detector aperture can also reduce the statistical noise, the pixel width is varied to guide the choice of detectors. From the profiles displayed in Fig. 6(a), it is difficult to separate the signal from the noise in the reconstructed image of undersized detector case, resulting in large error and deviation for the evaluation. At the same time, large detector width will degrade the spatial resolution as suggested by Eq. 1. So a 150μm-aperture cannot allow a resolution better than 100μm (see Fig. 6(b), (c) and Table 1) limited by its inherent measurement capability. Therefore, synthesizing the noise level and resolving ability, a pixel size of 100μm is suitable for the detector in this system.

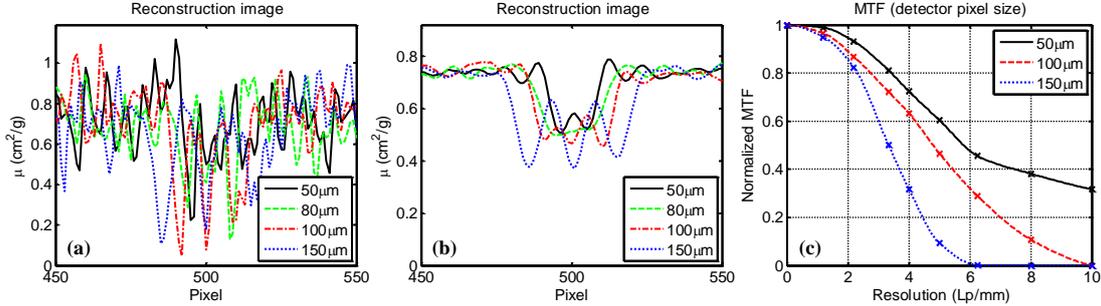

FIG. 6. Reconstructed results (central profiles) of different detector pixel sizes: (a) 50μm; (b) 150μm; (c) MTF image.

### 3.2 Practical requirements

As mentioned above, the crucial factors of a real laser-driven X-ray source that affect the CT performance, such as size fluctuation and position drift are unstable and difficult to predict, so the system characteristics are further discussed to assess the practical requirements on source parameters. The photon number per every DR image is set to be $2.8 \times 10^{10}$ to avoid the influence of additional statistical noises.

For source sizes range from 100μm to 1000μm, the resolutions derived from the MTF figure (Fig. 7(a)) are shown in Fig. 7(c). It is clear that the image contrast and the spatial resolution become worse when the source sizes gradually increase, in accordance with Eq. 1. A 100μm spatial resolution is accessible with a spot FWHM less than 500μm; otherwise the measurement precision will be substantially deteriorated.

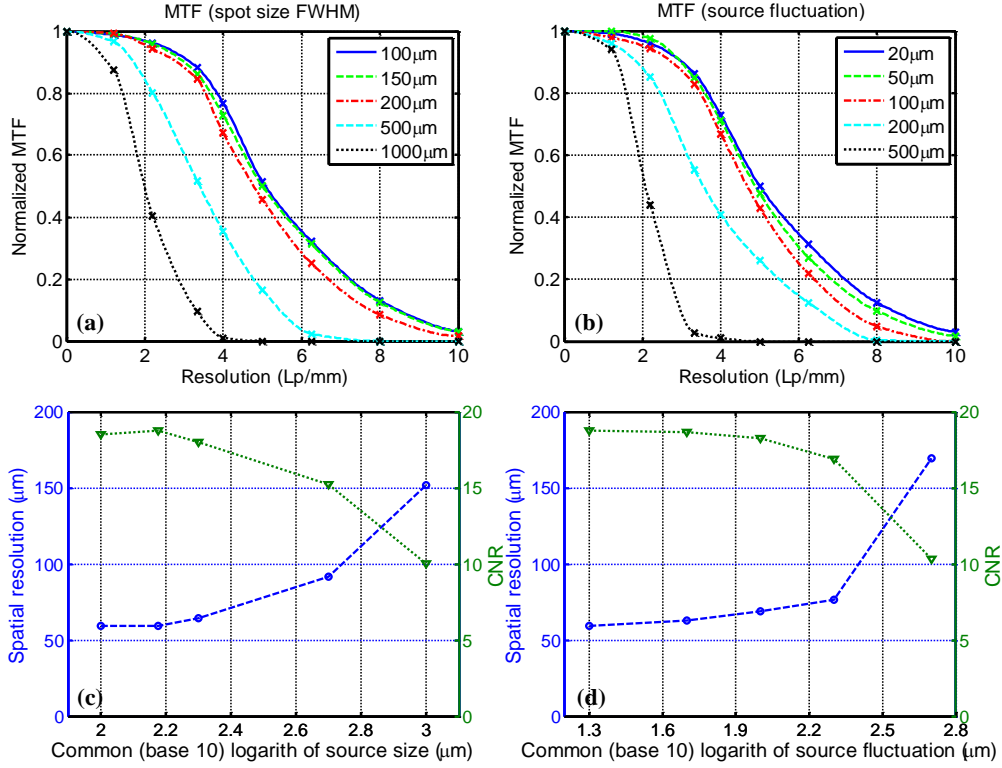

FIG. 7. MTF images and system performance of (a)(c) different source sizes and (b)(d) different source fluctuations.

Finally, the request of the CT system for actual instability is defined by changing the fluctuation range of source position and its size. From Fig. 7(b), we can find that the spatial resolution has a stronger dependence on the source fluctuation than its size due to the huge impact on the equivalent beam width, the rotation angle and the geometric position. Concluded from Fig. 7(d), the admissive random variation of source parameter is 200μm to ensure a resolution better than 100μm. Therefore, the improvement of laser performance and target design can benefit the stability of X-ray source and avoid adverse effects in the experiment.

Synthesized from the results above, an optimization solution of high-energy laser-based micro-CT can be summarized. To guarantee an acceptable image contrast and a spatial resolution better than 100μm, a series of parameters are required for current system: at least $1.4 \times 10^8$ photons/DR and 720 projection angles; an appropriate detector aperture of 100μm; an optimized magnification factor of 1.3; the FWHM of equivalent source spot within 500μm and the allowable tolerance of 200μm for spot size and position. The overall consideration of detector width, geometrical configuration, source size and statistical noise is necessary, and a properly chosen data set can considerably raise the performance level.

Additionally, a series of simulations with the optimized parameters using higher density material tungsten (19.35g/cm$^3$) are conducted to characterize the system capability. As illustrated in Fig. 8, the spatial resolution of reconstructed image is about 64.9μm with a CNR of 12.9, similar to the case of iron, proving the ability of this CT system to discriminate narrow crevices in complex and dense matters. This feature is of great interest for non-destructive testing (NDT) to examine the integrity and defect of industrial products.

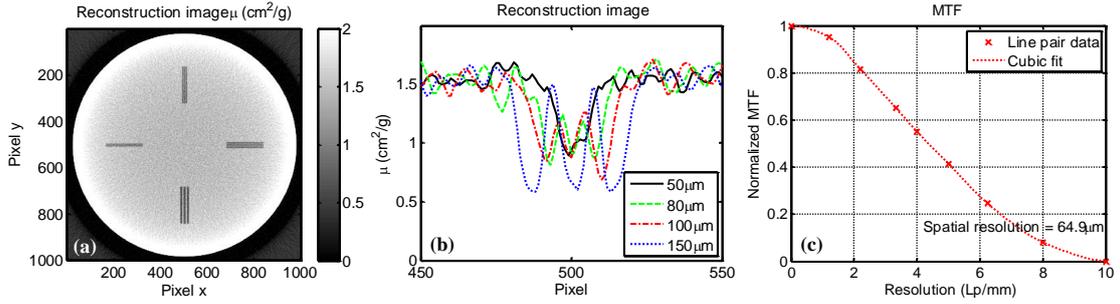

FIG. 8. Reconstructed results: (a) CT image, (b) central profiles and (c) MTF image of tungsten object.

According to our study, controlling the source fluctuation and enhancing the brilliance are main points to the progress of laser-driven X-ray source. And the optimization based on reachable condition can offer pointers to the design of actual pragmatic CT system to further improve the tomographic quality. After the conceptual study, an experiment has been accordingly designed and performed to generate a laser-based hard X-ray source for CT imaging. And a proof-of-principle demonstration has been successfully carried out with the micro-spot X-ray source from laser wakefield accelerator, which is presented and discussed in a separate paper [28]. As a result, we obtained the sinogram (Fig. 9(a)) and reconstructed distribution (Fig. 9(b)) of one cross-section of an object. A basic spatial resolution of about 100 micrometers has been achieved as shown in Fig. 9(c), which can be hopefully improved with the development of laser devices and target design. These results verify our conceptual design and optimization, laying the foundation for practically realizing the laser-plasma X-ray CT.

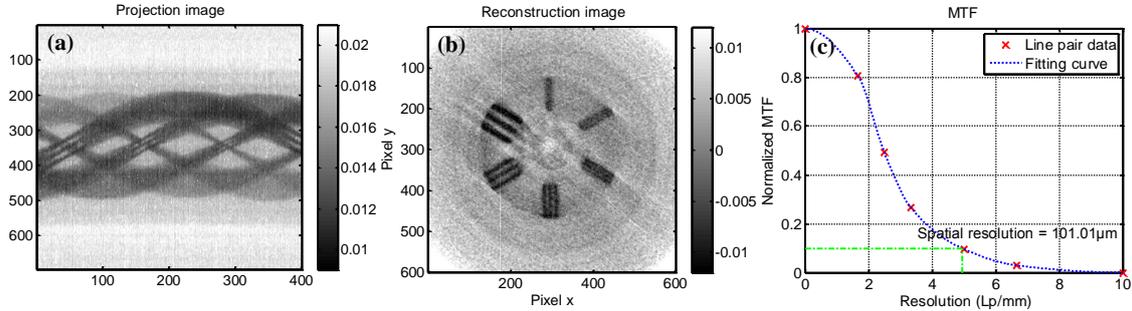

FIG. 9. Experimental result: (a) the sinogram (position within view vs. view angle); (b) Reconstructed CT image; (d) MTF image, in which the basic spatial resolution has reached ~100μm with 10% contrast standard.

## 4. Summary

To conclude, we propose a novel high-energy X-ray micro-CT based on laser-plasma interactions, and prove its practicability and feasibility by computational simulations. It is shown that a high-spatial-resolution (64.9μm) tomography of high-density objects (19.35g/cm$^3$) can be achieved by using typical laser-based hard X-ray source with optimized configurations. In the emulational models, the effects of realistic factors including parameter fluctuations, statistical noise and absorption efficiency of detector are basically introduced. And we evaluate the dependences of imaging quality on detector, source, scanning geometry and projection parameters. The tradeoff between spatial resolution and statistical noise is essential to be considered in the preliminary design. As a result, a consistent set of system parameters is developed to reasonably satisfy the performance requirements.

This concept provides a promising technical route to achieve high-resolution high-perspectivity CT by table-top laser devices which can simultaneously produce a very small X-ray spot with considerable energy and flux. It exhibits a large potential of the novel high-energy micro-CT in security check, precision inspection, material discrimination and many other applications [23].

**Acknowledgments**

This work is supported by the National Key R&D Program of China (No. 2016YFA0401100), the Science Challenge Program (Project No. TZ2016005, TZ2017005), the Science and Technology on Plasma Physics Laboratory at CAEP (No. 6142A04020102), and the Presidential Foundation of CAEP (Grant No. 201401017).